\documentclass[submission,copyright,creativecommons]{eptcs}
\providecommand{\event}{AFL 2014} 
\usepackage{breakurl}             

\title{A Simple Character String Proof\\
of the ``True but Unprovable'' Version\\
of G\"odel's First Incompleteness Theorem}
\author{Antti Valmari
\institute{Tampere University of Technology, Department of Mathematics\\
PO Box 553, FI-33101 Tampere, FINLAND}
\email{Antti.Valmari@tut.fi}
}
\def\titlerunning{Character String proof of First Incompleteness Theorem}
\def\authorrunning{Antti Valmari}

\usepackage{latexsym}
\usepackage{amssymb}

\newcommand{\ttt}{\texttt}
\newcommand{\ts}{\textsf}
\newcommand{\bs}{\char`\\}
\newcommand{\tl}{\char`~}
\newcommand{\nq}{\makebox[0pt][l]{/}=}
\newcommand{\defby}{\ $:\Leftrightarrow$\ }
\newcommand{\mb}{\mbox{}}
\newcommand{\mB}[1]{\makebox[#1][l]}
\newtheorem{theorem}{\textbf{Theorem}}{}

\begin{document}
\maketitle

\begin{abstract}
A rather easy yet rigorous proof of a version of G\"odel's first
incompleteness theorem is presented.
The version is ``each recursively enumerable theory of natural numbers with
$0$, $1$, $+$, $\cdot$, $=$, $\wedge$, $\neg$, and $\forall$ either proves a
false sentence or fails to prove a true sentence''.
The proof proceeds by first showing a similar result on theories of finite
character strings, and then transporting it to natural numbers, by using them
to model strings and their concatenation.
Proof systems are expressed via Turing machines that halt if and only if their
input string is a theorem.
This approach makes it possible to present all but one parts of the proof
rather briefly with simple and straightforward constructions.
The details require some care, but do not require significant background
knowledge.
The missing part is the widely known fact that Turing machines can perform
complicated computational tasks.
\end{abstract}


\section{Introduction}\label{S:intro}

Kurt G\"odel's first incompleteness theorem~\cite{God31} is certainly one of
the most important results in mathematical logic.
Together with an improvement by Barkley Rosser~\cite{Ros36}, the theorem says
that \emph{any recursive sufficiently strong theory of natural numbers either
proves a contradiction, or leaves both some sentence and its negation without
a proof}.
(We postpone discussion on G\"odel's original formulation to
Section~\ref{S:versions}, because it uses a concept that cannot be explained
briefly at this stage.
``Recursive'' and other background concepts are informally introduced in
Section~\ref{S:background}.)

More recently, the theorem has often been presented in the form
\emph{any recursively enumerable sufficiently expressive theory of natural
numbers either proves a sentence that does not hold or fails to prove a
sentence that does hold}.
This form is not equivalent to G\"odel's and Rosser's formulation.
In some sense it promises less and in some sense more.
However, it is easier to prove and perhaps also easier to understand.
It makes the assumption of sufficient expressiveness (explained in
Section~\ref{S:background}) instead of the stronger assumption of sufficient
strength (explained in Section~\ref{S:versions}).
This is the version discussed in the major part of this paper.
It is compared to G\"odel's and Rosser's formulation in
Section~\ref{S:versions}.

Both G\"odel's original proof and most of the modern expositions are long and
technical.
On the other hand, its overall strategy can be explained rather briefly and is
intuitively inspiring.
As a consequence, the proof is one of the most popularized ones.
We only mention here the excellent book by Douglas R.\
Hofstadter~\cite{Hof79}.
Unfortunately, to really grasp the result, the technicalities are necessary.

The goal of this paper is to present a \emph{rigorous} proof which, excluding
one detail, can be checked \emph{in full} by a reader with \emph{little
background} (but not necessarily with little effort).
We hope that our proof makes the result accessible to a wider audience than
before.
The skipped detail is the fact that some simple things can be computed by
so-called Turing machines.
Its rigorous proof would take many dull pages.
On the other hand, Turing machines have been very widely accepted as a
universal theoretical model of computation.
Therefore, as long as it is obvious that something could be programmed in a
modern programming language, it is common practice to skip the proof that a
Turing machine can compute it.

Our trick is to first prove that theories of finite character strings with
string literals, concatenation, and equality are incomplete.
Then we derive the incompleteness of natural number arithmetic as a corollary.
In this way, the main constructions of the proof are made using finite
character strings, while other proofs make them using natural numbers.
This makes our constructions much simpler and much more understandable.
The presentation of our proof in this paper is not remarkably short, but this
is partly due to the fact that it is very detailed.

Some background concepts are informally explained in
Section~\ref{S:background}.
Our language on finite character strings is defined in
Section~\ref{S:language}.
Not every character can be represented by itself in a string literal.
Therefore, an encoding of characters is needed.
Section~\ref{S:quotation} shows that the claim ``string $y$ is the sequence of
the encodings of the characters in string $x$'' can be formulated in the
language.
Computations of Turing machines are encoded in Section~\ref{S:machine}.
The incompleteness of theories of strings is shown in Section~\ref{S:strings},
and of theories of arithmetic in Section~\ref{S:arithmetic}.
Section~\ref{S:versions} compares the version of the theorem in this paper to
G\"odel's and Rosser's versions.
Discussion on related work and the conclusions are in
Section~\ref{S:conclusions}.

An earlier, not peer-reviewed version of this paper appeared as
arXiv:1402.7253v1.

\section{Informal Background}\label{S:background}

A \emph{recursive theory} consists of a language for formulating claims about
some domain of discourse, together with a recursive proof system.
In the case of G\"odel's theorem, the domain of discourse is the natural
numbers $0$, $1$, $2$, \ldots\ together with addition ($+$), multiplication
(variably denoted with $\times$, $\cdot$, \ttt{*}, or nothing such as in
$3x+1$), and equality ($=$) with their familiar properties.
A \emph{sentence} is a claim without input, formulated in the language.
For instance, ``3 is a prime number'' lacks input but ``$p$ is a prime
number'' has $p$ as input.
Of course, whether or not a claim can be formulated depends on the language.

When G\"odel published his theorem, the notion of ``recursive proof system''
had not yet developed into its modern form.
Indeed, instead of ``recursive'', he used a word that is usually translated as
``effective''.
G\"odel meant a mathematical reasoning system for proving sentences, where any
proof could be checked against a fixed set of straigthforward rules.
Proofs were checked by humans, but the requirement was that they could do that
in a mechanical fashion, without appealing to intuition on the meaning of
formulae.
This makes the proof system independent of the different insights that
different people might have.

Today, ``in a mechanical fashion'' means, in essence, ``with a computer that
has at least as much memory as needed''.
It suffices that there is a computer program that inputs a finite character
string and eventually halts if it is a valid proof, and otherwise runs
forever.
If such a program exists, then there also is a program that systematically
starts the former program on finite character strings one by one in increasing
length and executes them in parallel until a proof for the given sentence is
found.
(We will see in Section~\ref{S:arithmetic} how all finite character strings
can be scanned systematically.)
If the sentence has a proof, the program eventually finds it and halts;
otherwise it runs forever in a futile attempt to find a proof.

For mathematical analysis, computers and their programs are usually formalized
as \emph{Turing machines}.
We will introduce Turing machines in Section~\ref{S:machine}.

We will not need the assumption that a proof system resembles mathematical
reasoning systems.
Indeed, we will not need any other assumption than machine-checkability.
So we define a \emph{recursively enumerable proof system} as any Turing
machine $M$ that reads a finite character string and does or does not halt,
such that if the string is not a sentence, then $M$ does not halt.
If the string is a sentence, it is considered as proven if and only if $M$
halts.
A \emph{recursive proof system} adds to this the requirement of the existence
of another Turing machine that halts precisely on those inputs, on which $M$
does not halt.

Section~\ref{S:intro} assumed that the language for expressing claims about
natural numbers is sufficiently expressive.
It suffices that the language has constant symbols $0$ and $1$, an unbounded
supply of variable symbols, binary arithmetic operators $+$ and $\cdot$,
binary relation symbol $=$, binary logical operator $\wedge$ (that is,
``and''), unary logical operator $\neg$ (that is, ``not''), the so-called
universal quantifier $\forall$, and parentheses $($~and~$)$.
All symbols have their familiar syntactical rules and meanings.
The universal quantifier is used to formulate claims of the form $\forall n:
P(n)$ (that is, ``for every natural number $n$, $P(n)$ holds'').

For convenience, logical or $\vee$ can also be used without changing the
expressiveness of the language, because it can be built from $\wedge$ and
$\neg$, since $P \vee Q$ is logically equivalent to $\neg(\neg P \wedge \neg
Q)$.
Also logical implication $\rightarrow$, existential quantifier $\exists$,
inequality $\neq$, less than $<$, and all familiar numeric constants $2$, $3$,
\ldots, $9$, $10$, $11$, \ldots\ can be used, because $P \rightarrow Q$ is
equivalent to $\neg(P \wedge \neg Q)$, $\exists x: P(x)$ is equivalent to
$\neg\forall x: \neg P(x)$, $x \neq y$ is equivalent to $\neg(x=y)$, $x < y$
is equivalent to $(\exists z: x+z+1=y)$, and any such numeric constant has the
same value as some expression of the form $(1+1+\ldots+1)$.

\section{A First-Order Language on Finite Character Strings}\label{S:language}

In this section we define our language for expressing claims about
\emph{finite character strings}, that is, finite sequences of characters.
We use\ttt{ this font }when writing in that language.
To make it explicit where a string in that language ends and ordinary text
continues, we put white space on both sides of the string even if normal
writing rules of English would tell us not to do so.
So we write ``characters\ttt{ a }, \ldots,\ttt{ z }are'' instead of
``characters \ttt{a}, \ldots, \ttt{z} are''.

The language uses the following characters:
\begin{quote}
\ttt{%
a b c d e f g h i j k l m n o p q r s t u v w x y z\\
0 1 2 3 4 5 6 7 8 9 " \bs\ = \nq\ ( ) \tl\ \& | - > A E : + * < }
\end{quote}
We chose this set of characters for convenience.
Any finite set containing at least two characters could have been used, at the
expense of a somewhat more complicated proof.
This set facilitates the use of familiar notation for many things.
The fact that its size 53 is a prime number will be exploited in
Section~\ref{S:arithmetic}.
The characters\ttt{ a }, \ldots,\ttt{ z }are \emph{lower case letters},
and\ttt{ 0 }, \ldots,\ttt{ 9 }are \emph{digits}.

A \emph{finite character string} or just \emph{string} is any finite sequence
of characters.

An \emph{encoded character} is\ttt{ \bs0 },\ttt{ \bs1 }, or any other
character than\ttt{ " }and\ttt{ \bs\ }.
The encoded character\ttt{ \bs0 }denotes the character\ttt{ \bs\ },\ttt{
\bs1\ }denotes\ttt{ " }, and each remaining encoded character denotes itself.
A \emph{string literal} is any sequence of characters of the form\ttt{
"$\alpha$" }, where $\alpha$ is any finite sequence of encoded characters.
It denotes the corresponding sequence of (unencoded) characters.
For instance,\ttt{ "" }denotes the empty string and\ttt{
"backslash=\bs1\bs0\bs1" }denotes the string\ttt{ backslash="\bs" }.
The purpose of encoding is to facilitate the writing of\ttt{ " }inside a
string literal, without causing confusion with the\ttt{ " }that marks the end
of the literal.

A \emph{variable} is any string that starts with a lower case letter and then
consists of zero or more digits.
For instance,\ttt{ a },\ttt{ x0 }, and\ttt{ y365 }are variables but\ttt{ 49
}and\ttt{ cnt }are not.
The value of a variable is a string.
We say that the variable \emph{contains} the string.

A \emph{term} is any non-empty finite sequence of variables and/or string
literals.
It denotes the concatenation of the strings that the variables contain and/or
string literals denote.
For instance,\ttt{ "theorem" },\ttt{ "theo""rem" }and\ttt{ "the""""o""rem"
}denote the same string\ttt{ theorem }.
If the variable\ttt{ x }contains the string\ttt{ or }, then also\ttt{
"the"x"em" }denotes\ttt{ theorem }.

An \emph{atomic proposition} is any string of the form\ttt{ $t$=$u$ }or of the
form\ttt{ $t$\nq$u$ }, where $t$ and $u$ are terms.
The first one expresses the claim that the strings denoted by $t$ and $u$ are
the same string, and the second one expresses the opposite claim.
So\ttt{ "theorem"="theo""rem" }is a true atomic proposition, and\ttt{
"theorem"\nq"theo""rem" }is not.
Indeed,\ttt{ $t$\nq$u$ }expresses the same claim as\ttt{ \tl$t$=$u$ },
where\ttt{ \tl\ }is introduced soon.

A \emph{formula} is either an atomic proposition or any string of the
following forms, where $\pi$ and $\rho$ are formulae and $x$ is a variable:%
\ttt{~($\pi$) },
\ttt{ \tl$\pi$ },
\ttt{ $\pi$\&$\rho$ },
\ttt{ $\pi$|$\rho$ },
\ttt{ $\pi$->$\rho$ },
\ttt{ A$x$:$\pi$ }, and
\ttt{ E$x$:$\pi$ }.
The parentheses\ttt{ ( }and\ttt{ ) }are used like in everyday mathematics, to
force the intended interpretation.
In the absence of parentheses, formulae are interpreted according to the
following precedences: concatenation has the highest precedence, then\ttt{ =
},\ttt{ \tl\ },\ttt{ \& },\ttt{ | },\ttt{ -> }, and\ttt{ :~}in this order.
For instance,\ttt{ \tl{}b=c|c="hello"\&a=bc }denotes the same as\ttt{
(\tl(b=c))|((c="hello")\&(a=bc)) }, and\\\ttt{ Ax:\tl x="8"|x="8"->"u"="u"
}denotes the same as\ttt{ Ax:(((\tl(x="8"))|(x="8"))->("u"="u")) }.
All operators associate to the left, so\ttt{ $\pi$->$\rho$->$\sigma$ }means
the same as\ttt{ ($\pi$->$\rho$)->$\sigma$ }.

The formulae express the following claims:
\begin{center}
\begin{tabular}{ll}
\ttt{($\pi$)} & expresses the same claim as $\pi$,\\
\ttt{\tl$\pi$} & expresses that $\pi$ does not hold,\\
\ttt{$\pi$\&$\rho$} & expresses that $\pi$ and $\rho$ hold,\\
\ttt{$\pi$|$\rho$} & expresses that $\pi$ holds or $\rho$ holds or both
hold,\\
\ttt{$\pi$->$\rho$} & expresses that if $\pi$ holds, then also $\rho$ holds,\\
\ttt{A$x$:$\pi$} & expresses that for any string $x$, $\pi$ holds, and\\
\ttt{E$x$:$\pi$} & expresses that there is a string $x$ such that $\pi$
holds.
\end{tabular}
\end{center}

To improve readability, we often add spaces into a formula, like\ttt{
Ax:~x\nq"8" | x="8" }.
We may also split a formula onto many lines.

A \emph{first-order language} is any language whose formulae are built from
atomic propositions like above.
The constants or literals, terms, and atomic propositions of a first-order
language may be chosen as appropriate to the domain of discourse.
A variable of a first-order language may only contain a value in the domain of
discourse, while a variable of a higher-order language may be used more
flexibly.
When talking about first-order languages in general, we use the symbols
$\neg$, $\forall$, and so on, and when talking about a particular first-order
language specified in this paper, we use\ttt{ \tl~},\ttt{ A }, and so on.

\newcommand{\Char}[1]{\ts{Char(}#1\ts{)}}
We will need long formulae.
To simplify reading them, we introduce abbreviations.
The first abbreviation claims that variable\ttt{ a }contains a character, that
is, a string of length precisely one.
The formula consists of\ttt{ a="$x$" }for each encoded character $x$,
separated by\ttt{ | }and surrounded by\ttt{ ( }and\ttt{ ) }.
We do not write it in full but instead write $\cdots$ to denote the missing
part.
\begin{quote}
\ttt{\Char{a} \defby\ ( a="a" | a="b" | $\cdots$ | a="<" )}
\end{quote}
The abbreviation was written for variable\ttt{ a }, but clearly a similar
abbreviation can be written for any variable.
So we may use the abbreviations\ttt{ \Char{b} },\ttt{ \Char{g75} }, and so on.

\newcommand{\Sb}[2]{\ts{Sb(}#1\ts{, }#2\ts{)}}
The next abbreviation claims that\ttt{ x }is a substring of\ttt{ y }.
That is, there are strings\ttt{ u }and\ttt{ v }such that string\ttt{ y }is the
same as string\ttt{ u }followed by string\ttt{ x }followed by string\ttt{ v }.
\begin{quote}
\ttt{\Sb{x}{y} \defby\ ( Eu:Ev:~y=uxv )}
\end{quote}
When this abbreviation is used with\ttt{ u }in the place of\ttt{ x }, some
other variable has to be used instead of\ttt{ u }on the right hand side.
That is,\ttt{ \Sb{u}{y} }does not abbreviate\ttt{ ( Eu:Ev:~y=uuv ) }but, for
instance,\ttt{ (~Ez:Ev:~y=zuv ) }.
The incorrect interpretation\ttt{ ( Eu:Ev:~y=uuv ) }contains a \emph{name
clash}, that is, the variable\ttt{ x }that is distinct from\ttt{ u }in\ttt{
y=uxv }in the definition of\ttt{ \Sb{x}{y} }, became the same variable as\ttt{
u }.
In general, when interpreting an abbreviation containing a subformula of the
form\ttt{ A$x$:$\pi$ }or\ttt{ E$x$:$\pi$ }, it may be necessary to replace $x$
by some other variable, to avoid name clashes.
Further information on this issue can be found in textbooks on logic, in
passages that discuss ``bound'' and ``free'' variables.

Please keep in mind that abbreviations are not part of our language.
They are only a tool for compactly referring to certain formulae that are too
long to be written in full.
Each string that uses abbreviations denotes the string that is obtained by
replacing the abbreviations by their definitions, changing variable names in
the definitions as necessary to avoid name clashes.

\section{A Formula Expressing the Encoding of Characters}\label{S:quotation}
\newcommand{\Q}[2]{\ts{Q(}#1\ts{,~}#2\ts{)}}
\newcommand{\Qp}[2]{\ts{\.Q(}#1\ts{,~}#2\ts{)}}

In this section we show that a formula\ttt{ \Q{x}{y} }can be written that
claims that\ttt{ y }is the encoding of\ttt{ x~}, that is,\ttt{ y }is obtained
by replacing\ttt{ \bs0 }for each\ttt{ \bs\ }and\ttt{ \bs1 }for each\ttt{ "
}in\ttt{ x }.
We start with a formula claiming that\ttt{ y }is obtained by replacing\ttt{ v
}for one instance of\ttt{ u }inside\ttt{ x~}.
\newcommand{\RepOne}[4]{\ts{RepOne(}#1\ts{, }#2\ts{, }#3\ts{, }#4\ts{)}}
\begin{quote}
\ttt{\RepOne{x}{u}{v}{y} \defby\ ( Ee:Ef:~x=euf \& y=evf )}
\end{quote}

The next formula claims that, under certain assumptions mentioned below,\ttt{
y }is obtained by replacing\ttt{ v }for every instance of\ttt{ u
}inside\ttt{ x }.
It converts\ttt{ x }to\ttt{ y }by making the replacements one by one.
It assumes that\ttt{ v }has no characters in common with\ttt{ u }, so that no
fake instances of\ttt{ u }can occur inside or overlapping\ttt{ v }.
Furthermore, it assumes that different instances of\ttt{ u }in\ttt{ x }do not
overlap, so that the result is independent of the order in which the instances
are chosen for replacement.
It also assumes that\ttt{ p }(for punctuation) is a string that does not occur
inside\ttt{ x },\ttt{ y }, or any intermediate result.
Furthermore,\ttt{ p }cannot overlap with itself.
We will later see how\ttt{ p }is constructed.

The sequence of replacements is represented by\ttt{ s }as a sequence of the
form\ttt{ p$x_1$p$x_2$p$\cdots$p$x_n$p }, where $x_1=\ttt{x}$~,
$x_n=\ttt{y}$~, and $x_2$, \ldots, $x_{n-1}$ are the intermediate results.
The requirements on\ttt{ p }guarantee that\ttt{ s }can be decomposed into this
form in precisely one way.
The parts\ttt{ (Et:~s=pxpt) }and\ttt{ (Et:~s=tpyp) }guarantee that
$x_1=\ttt{x}$ and $x_n=\ttt{y}$~.
Thanks to\ttt{ \tl\Sb{u}{y} }, \emph{every} instance of\ttt{ u }is replaced.
The rest of the formula picks each $x_i$ other than the last and claims that
$x_{i+1}$ is obtained from it by making one replacement.
The $x_i$ is represented by\ttt{ h }and $x_{i+1}$ by\ttt{ k }.
They are distinguished by not containing\ttt{ p }, being preceded by\ttt{ p },
being separated from each other by\ttt{ p }, and being succeeded by\ttt{ p }.

\newcommand{\RepAll}[5]{\ts{RepAll(}#1\ts{, }#2\ts{, }#3\ts{, }#4\ts{,
}#5\ts{)}}
\begin{quote}
\ttt{\RepAll{x}{u}{v}{y}{p} \defby\ ( Es:\\
\mb\hspace{1cm}~ (Et:~s=pxpt) \& (Et:~s=tpyp) \& \tl\Sb{u}{y}\\
\mb\hspace{1cm}\& Ah:Ak:~( \Sb{phpkp}{s} \& \tl\Sb{p}{h} \& \tl\Sb{p}{k} )
-> \RepOne{h}{u}{v}{k}\\
)}
\end{quote}

To emphasize that abbreviations \emph{are not} but \emph{stand for} strings in
our language, and that the strings they stand for are often not easy to
comprehend, we now show the string that\ttt{ \RepAll{x}{u}{v}{y}{p} }stands
for.
The real string is too long to be shown on one line, so we split it on two
lines.
\begin{quote}
\ttt{(Es:(Et:s=pxpt)\&(Et:s=tpyp)\&\tl
(Ez:Ev:y=zuv)\&Ah:Ak:((Eu:Ev:s=uphpkpv)\&\\
\tl(Eu:Ev:h=upv)\&\tl(Eu:Ev:k=upv))->(Ee:Ef:h=euf\&k=evf))}
\end{quote}

To obtain the punctuation string\ttt{ p }, we first make variable\ttt{ q
}contain some sequence of \ttt{:}-characters that does not occur inside\ttt{ x
}.
Such a string exists, because the string consisting of $n+1$
\ttt{:}-characters meets the requirements, when $n$ is the length of\ttt{ x }.

\newcommand{\Punct}[2]{\ts{Punct(}#1\ts{, }#2\ts{)}}
\begin{quote}
\ttt{\Punct{x}{q} \defby\ ( ( Aa:~\Sb{a}{q} \& \Char{a} -> a=":" ) \&
\tl\Sb{q}{x} )}
\end{quote}

The\ttt{ p }used above is obtained as\ttt{ "+"q }, that is, by adding a
\ttt{+}-character to the front of the sequence of \ttt{:}-characters in
variable\ttt{ q }.
So the value of\ttt{ p }is\ttt{ +:::::~}or some similar sequence with a
different number of \ttt{:}-characters.
It clearly neither overlaps with itself nor occurs within\ttt{ x }.

To guarantee that\ttt{ u }and\ttt{ v }do not have characters in common, we
first convert each instance of\ttt{ \bs\ }to\ttt{ "*"q }, that is, to some
string of the form\ttt{ *:::$\cdots$:~}that does not occur inside\ttt{ x }.
Then each\ttt{ *:::$\cdots$:~}is converted to\ttt{ \bs0 }, then each\ttt{ "
}to\ttt{ *:::$\cdots$:~}, and finally each\ttt{ *:::$\cdots$:~}to\ttt{ \bs1 }.
In the first conversion,\ttt{ u }consists of a single character, so different
instances of\ttt{ u }do not overlap.
The same holds for the third conversion.
In the second and fourth conversion,\ttt{ u }is\ttt{ *:::$\cdots$:~}, which
clearly cannot overlap with itself.
Furthermore,\ttt{ +:::$\cdots$:~}does not overlap and is not inside\ttt{
*:::$\cdots$:~}, so\ttt{ p }cannot occur in\ttt{ y }or any intermediate
result.

We are now ready to write\ttt{ \Q{x}{y} }.
In it, the value\ttt{ \bs\ }is represented by the string literal\ttt{ "\bs0"
},\ttt{ \bs1\ }by\ttt{ "\bs01" }, and so on.  
\begin{quote}
\ttt{\mB{25mm}{\Q{x}{y} \defby}( Eq:~~\Punct{x}{q}\\
\mB{25mm}{}\& Ex1:~\RepAll{ x}{"\bs0"}{ "*"q}{x1}{"+"q}\\
\mB{25mm}{}\& Ex2:~\RepAll{x1}{"*"q}{"\bs00"}{x2}{"+"q}\\
\mB{25mm}{}\& Ex3:~\RepAll{x2}{"\bs1"}{ "*"q}{x3}{"+"q}\\
\mB{25mm}{}\& ~~~~~\RepAll{x3}{"*"q}{"\bs01"}{ y}{"+"q}{ )}}
\end{quote}

\section{Encoding Turing Machine Computations}\label{S:machine}

\newcommand{\Pvble}[1]{\ts{Pvble(}#1\ts{)}}
\newcommand{\Pvblep}[1]{\ts{\.Pvble(}#1\ts{)}}
Turing machines are a formal model of computation.
In this section we show that for each Turing machine, there is a formula\ttt{
\Pvble{x} }that yields true if and only if the machine eventually halts,
given\ttt{ x }as the input.
We call it\ttt{ \Pvble{x} }, because the Turing machine is thought to
represent some proof system such that it halts if and only if\ttt{ x }can be
proven.

Details of the definition of Turing machines vary in the literature.
To start our definition, we introduce a new symbol $\sqcup$, called
\emph{blank}.
Let \emph{b-strings} be defined similarly to strings, but they may also
contain blanks.
So our Turing machines use $54$ symbols: $53$ characters and the blank.
A Turing machine consists of a \emph{control unit}, a \emph{read/write head},
and a \emph{tape} that consists of an infinite number of \emph{cells} in both
directions.
Each cell on the tape may contain any character or $\sqcup$.
When we say that some part of the tape is blank, we mean that each cell in it
contains $\sqcup$.
At any instant of time, the read/write head is on some cell of the tape.
During a computation step, the read/write head rewrites the content of the
cell and then possibly moves to the previous or the next cell, as dictated by
the control unit and the contents of the cell before the step.

The control unit consists of \emph{states} and \emph{rules}.
The states are numbered from $0$ to $r$, for some positive integer $r$.
State $0$ is called the \emph{final state}.
There are $54r$ rules, one for each state $q$ other than $0$ and for each
character $c$ and $\sqcup$.
A rule is of the form $(c,q) \mapsto (c',q',d)$, where $c'$ is any character
or $\sqcup$, $q'$ is any state, and $d$ is either\ttt{ L },\ttt{ R }, or\ttt{
N }.
The meaning of the rule is that if the control unit is in state $q$ and the
tape cell under the read/write head contains $c$, then the machine writes $c'$
on the cell, moves the read/write head one cell to the left or right or does
not move it, and the control unit enters its state $q'$.
If the control unit enters state $0$, then computation halts.

Initially, the tape contains a finite sequence of characters, written
somewhere on the tape.
This finite sequence is the input to the machine.
The rest of the tape is initially blank.
Initially, the read/write head is on the first input character (or just
anywhere, if the input is empty), and the control unit is in state $1$.

At any instant of time, let the \emph{right b-string} mean the content of the
cell under the read/write head, the content of the next cell to the right, and
so on, up to and including the last character on the tape.
If the cell under the read/write head and all cells to the right are blank,
then the right b-string is empty.
So the last symbol of a non-empty right b-string is always different from
$\sqcup$.
Let the \emph{left b-string} be defined similarly, but starting at the cell
immediately to the left of the read/write head, and proceeding to the left
until the first character on the tape is taken.
Again, the left b-string may be empty, and if it is not, then its last symbol
is not $\sqcup$.
The contents of the tape as a whole are an infinite sequence of blanks
extending to the left, then the left b-string reversed, then the right
b-string, and then an infinite sequence of blanks extending to the right.
Initially, the left b-string is empty and the right b-string contains the
input.

A halting computation corresponds to a sequence $(\lambda_0,q_0,\rho_0)$,
$(\lambda_1,q_1,\rho_1)$, \ldots, $(\lambda_n,q_n,\rho_n)$, where $\lambda_0$
is the empty string, $q_0=1$, $\rho_0$ is the input string, $q_n=0$, $q_i \neq
0$ when $0 \leq i < n$, and each $(\lambda_i,q_i,\rho_i)$ for $1 \leq i \leq
n$ is obtained from $(\lambda_{i-1},q_{i-1},\rho_{i-1})$ as follows.
Here $\lambda_i$ is the left b-string and $\rho_i$ is the right b-string after
$i$ computation steps.
Let $c=\sqcup$ if $\rho_{i-1}$ is empty, and otherwise let $c$ be the first
symbol of $\rho_{i-1}$.
There is a unique rule of the form $(c,q_{i-1}) \mapsto (c',q',d)$.
We have $q_i=q'$.
The b-strings $\lambda_i$ and $\rho_i$ are obtained by replacing $c'$ for the
first symbol of $\rho_{i-1}$, with special treatment of the case that
$\rho_{i-1}$ is empty or $c' = \sqcup$; and then possibly moving the first
symbol of the resulting b-string to the front of $\lambda_{i-1}$, or moving
a symbol in the opposite direction, again with some special cases.
The special cases are discussed in more detail later in this section.
The moving of a symbol from the right b-string to the left b-string models the
movement of the read/write head one cell to the right, and the moving of a
symbol in the opposite direction models the movement of the read/write head
one cell to the left.

We want to model this sequence in our language on strings.
The states $q_i$ are represented simply by writing their numbers using the
digits\ttt{ 0 },\ttt{ 1 }, \ldots,\ttt{ 9 }in the usual way.
That is, state number $32768$ is represented by\ttt{ 32768 }.
The b-strings $\lambda_i$ and $\rho_i$ are more difficult, because they may
contain blanks, but there is no blank character in our language.
So we represent $\sqcup$ with\ttt{ \bs2 }and\ttt{ \bs\ }with\ttt{ \bs0 }.
To simplify later constructions by remaining systematic with the encoding in
Section~\ref{S:language}, we also represent\ttt{ " }with\ttt{ \bs1 }.
All other characters represent themselves.
To summarize,\ttt{ \bs\ },\ttt{ " }, and $\sqcup$ on the tape of the Turing
machine are represented in $\lambda_i$ and $\rho_i$ by the values\ttt{ \bs0
},\ttt{ \bs1 }, and\ttt{ \bs2 }, whose string literal representations are\ttt{
"\bs00" },\ttt{ "\bs01" }, and\ttt{ "\bs02"~}.
In this sense,\ttt{ \bs\ }and\ttt{ " }become doubly encoded. 

So we define an \emph{encoded symbol} as\ttt{ \bs0 },\ttt{ \bs1 },\ttt{ \bs2
}, or any other character than\ttt{ \bs\ }and\ttt{ " }.
We need not (and could not) say that $\sqcup$ is not an encoded symbol,
because $\sqcup$ is not a character at all.

\newcommand{\EChar}[1]{\ts{EChar(}#1\ts{)}}
\begin{quote}
\ttt{\EChar{e} \defby\\
\mb\hspace{1cm}( e="\bs00" | e="\bs01" | e="\bs02" | \Char{e} \&
e\nq"\bs0" \& e\nq"\bs1" )}
\end{quote}

The next formula expresses that\ttt{ y }is obtained from\ttt{ x }by
replacing\ttt{ e }for its first encoded symbol, with special treatment of the
empty strings and the blank.
If\ttt{ x }consists of at most one encoded symbol,\ttt{ x }as a whole is
overwritten.
The result is the empty string if\ttt{ e }is the encoded blank, and otherwise
the result is\ttt{ e }.
If\ttt{ x }consists of more than one encoded symbols, ordinary replacement
occurs.

\newcommand{\Write}[3]{\ts{Write(}#1\ts{, }#2\ts{, }#3\ts{)}}
\begin{quote}
\ttt{\Write{x}{e}{y} \defby\ (\\
\mb\hspace{1cm}~ ( x="" | \EChar{x} ) \& ( e="\bs02" \& y="" | e\nq"\bs02" \&
y=e )\\
\mb\hspace{1cm}| ( Ef:Ez:~x=fz \& \EChar{f} \& z\nq"" \& y=ez )\\
)}
\end{quote}

Let the encoded form of the left b-string be called \emph{left string}, and
similarly with the right b-string.
The next formula expresses the removal of the first encoded symbol from one
string and its addition to the front of another string, again with special
treatment of the empty strings and the blank.
The variables\ttt{ f1 }and\ttt{ f2 }contain the values of the from-string
before and after the operation, and\ttt{ t1 }and\ttt{ t2 }contain the
to-string.
The encoded blank\ttt{ \bs2\ }is never added to the front of an empty
to-string, to maintain the rule that the b-strings never end with the blank.
If the from-string is empty, then the operation behaves as if the encoded
blank were extracted from it.

\newcommand{\Move}[4]{\ts{Move(}#1\ts{, }#2\ts{, }#3\ts{, }#4\ts{)}}
\begin{quote}
\ttt{\Move{f1}{t1}{f2}{t2} \defby\ (\\
\mb\hspace{1cm}~~( f1="" \& t1="" \& f2="" \& t2="" )\\
\mb\hspace{1cm}| ( f1="" \& t1\nq"" \& f2="" \& t2="\bs02"t1 )\\
\mb\hspace{1cm}| ( f1="\bs02"f2 \& t1="" \& t2="" )\\
\mb\hspace{1cm}| ( Ee:~\EChar{e} \& f1=ef2 \& (e\nq"\bs02" | t1\nq"") \&
t2=et1 )\\
)}
\end{quote}

Next we introduce a formula for each rule $(c,q) \mapsto (c',q',d)$.
Let $\dot c=\ttt{\bs0}$~, if $c=\ttt{\bs}$~; $\dot c=\ttt{\bs1}$~, if
$c=\ttt{"}$~; $\dot c=\ttt{\bs2}$~, if $c=\sqcup$~; and otherwise $\dot c=c$.
Let $\ddot c=\ttt{\bs00}$~, if $c=\ttt{\bs}$~; $\ddot c=\ttt{\bs01}$~, if
$c=\ttt{"}$~; $\ddot c=\ttt{\bs02}$~, if $c=\sqcup$~; and otherwise $\ddot
c=c$.
We define $\dot c'$ and $\ddot c'$ similarly.
Let $\dot q$ denote $q$ written using\ttt{ 0 },\ttt{ 1 }, \ldots,\ttt{ 9 }in
the usual way, and similarly with $\dot q'$.

We consider first the case where $d=\ttt{N}$.
The first part of the formula checks that the rule triggers, that is, the
current state is $q$ and the symbol under the read/write head is $c$, taking
into accout the possibility that the right string is empty.
The second part of the formula gives the state of the control unit, the right
string, and the left string new values as dictated by the rule.
\begin{quote}
\ttt{\ts{Rule$_{c,q}^{c',q',\ttt{N}}$(}l1\ts{, }r1\ts{, }q1\ts{, }l2\ts{,
}r2\ts{, }q2\ts{)} \defby\ (\\
\mb\hspace{1cm}~ q1="$\dot q$" \& ( r1="" \& "$\ddot c$"="\bs02" |
Ex:~r1="$\ddot c$"x )\\
\mb\hspace{1cm}\& q2="$\dot q'$" \& \Write{r1}{"$\ddot c'$"}{r2} \& l2=l1\\
)}
\end{quote}
Rules with $d=\ttt{R}$ or $d=\ttt{L}$ are similar, but the moving of the
read/write head is also represented.
\begin{quote}
\ttt{\ts{Rule$_{c,q}^{c',q',\ttt{R}}$(}l1\ts{, }r1\ts{, }q1\ts{, }l2\ts{,
}r2\ts{, }q2\ts{)} \defby\ (\\
\mb\hspace{1cm}~ q1="$\dot q$" \& ( r1="" \& "$\ddot c$"="\bs02" |
Ex:~r1="$\ddot c$"x )\\
\mb\hspace{1cm}\& q2="$\dot q'$" \& Er:~\Write{r1}{"$\ddot c'$"}{r} \&
\Move{r}{l1}{r2}{l2}\\
)}
\end{quote}
\begin{quote}
\ttt{\ts{Rule$_{c,q}^{c',q',\ttt{L}}$(}l1\ts{, }r1\ts{, }q1\ts{, }l2\ts{,
}r2\ts{, }q2\ts{)} \defby\ (\\
\mb\hspace{1cm}~ q1="$\dot q$" \& ( r1="" \& "$\ddot c$"="\bs02" |
Ex:~r1="$\ddot c$"x )\\
\mb\hspace{1cm}\& q2="$\dot q'$" \& Er:~\Write{r1}{"$\ddot c'$"}{r} \&
\Move{l1}{r}{l2}{r2}\\
)}
\end{quote}

Let $\ddot\lambda_i$ and $\ddot\rho_i$ be obtained from $\lambda_i$ and
$\rho_i$ by replacing each symbol $c$ in them with $\ddot c$.
The computation of the Turing machine is represented as a string\ttt{ c }of
the form
\begin{quote}
\ttt{ "$\ddot\lambda_0$"$\ddot\rho_0$\bs3$\dot
q_0$"$\ddot\lambda_1$"$\ddot\rho_1$\bs3$\dot q_1$"%
$\cdots$"$\ddot\lambda_n$"$\ddot\rho_n$\bs3$\dot q_n$" }.
\end{quote}
Because the $\dot q_i$ consist of just digits and the $\ddot \lambda_i$ and
$\ddot \rho_i$ have been encoded,\ttt{ " }and\ttt{ \bs3\ }cannot occur inside
them.
So they can be used for separating the $\ddot\lambda_i$, $\ddot\rho_i$, and
$\dot q_i$ from each other.

We are ready to write the formula that claims that the Turing machine halts on
input\ttt{ x }.
It says that there is a sequence\ttt{ c }that models the computation.
First,\ttt{ c }starts with the empty left string, the encoded input string as
the right string, and $1$ as the state.
Second,\ttt{ c }ends with $0$ as the state.
Finally, each\ttt{ "$\ddot\lambda_i$"$\ddot\rho_i$\bs3$\dot
q_i$"$\ddot\lambda_{i+1}$"$\ddot\rho_{i+1}$\bs3$\dot q_{i+1}$" }satisfies some
rule.
That\ttt{ q1 }and\ttt{ q2 }do not pick more from\ttt{ c }than they should
follows from the fact that the rules check that they consist of digits only.

\begin{quote}
\ttt{\Pvble{x} \defby\ ( Ec:\\
\mb\hspace{1cm}~ ( Et:Ey:~\Q{x}{y} \& c="\bs1\bs1"y"\bs031\bs1"t ) \& (
Et:~c=t"\bs030\bs1" )\\
\mb\hspace{1cm}\& ( Al1:Ar1:Aq1:~Al2:Ar2:Aq2:\\
\mb\hspace{1cm}~ ~ ~ ~ ~\tl\Sb{"\bs1"}{l1} \& \tl\Sb{"\bs1"}{r1} \&
\tl\Sb{"\bs1"}{l2} \& \tl\Sb{"\bs1"}{r2}\\
\mb\hspace{1cm} ~ ~ ~ \&
\Sb{"\bs1"l1"\bs1"r1"\bs03"q1"\bs1"l2"\bs1"r2"\bs03"q2"\bs1"}{c}\\
\mb\hspace{1cm} ~ ~-> ( \mB{15mm}{}\mB{12mm}{\ts{Rule$_1$}}\ts{(}l1\ts{,
}r1\ts{, }q1\ts{, }l2\ts{, }r2\ts{, }q2\ts{)}\\
\mb\hspace{1cm} ~ ~ ~ ~ \mB{15mm}{| $\cdots$
|}\mB{12mm}{\ts{Rule$_{54r}$}}\ts{(}l1\ts{, }r1\ts{, }q1\ts{, }l2\ts{,
}r2\ts{, }q2\ts{)} ) )\\
)}
\end{quote}

\section{Incompleteness of Theories of Finite Character
Strings}\label{S:strings}

In this section we prove that any recursively enumerable proof system for our
language on strings either fails to prove some true sentence, or proves some
false sentence.

Please remember that\ttt{ \Q{x}{y} }and\ttt{ \Pvble{x"\bs1"y"\bs1"} }are
abbreviations used in this paper to improve readability, and not as such
strings in our language.
They stand for some strings in our language that are too long to be written
explicitly in this paper.
Each of these two long strings has a corresponding encoded string, which is
obtained by replacing\ttt{ \bs0\ }for each\ttt{ \bs\ }and\ttt{ \bs1\ }for
each\ttt{ " }.
We denote them with\ttt{ \Qp{x}{y} }and\ttt{ \Pvblep{x"\bs1"y"\bs1"} }.
Also remember that\ttt{ \Q{x}{y} }claims that\ttt{ y }is the encoded form
of\ttt{ x }.
Therefore,
\begin{quote}
\ttt{\Q{ \Q{x}{y}}{ \Qp{x}{y} }}\hfill and\hfill\ttt{\Q{
\Pvble{x"\bs1"y"\bs1"}}{ \Pvblep{x"\bs1"y"\bs1"} }}\hfill hold.\hfill\mb
\end{quote}

We can now write G\"odel's famous self-referential sentence in our framework
as follows.

\begin{quote}
\ttt{ Ex:Ey:~\Q{x}{y} \& \tl\Pvble{x"\bs1"y"\bs1"} \& x=\\
"Ex:Ey:~\Qp{x}{y} \& \tl\Pvblep{x"\bs1"y"\bs1"} \& x="}
\end{quote}
Let $\alpha$ be any string and $\beta$ be its encoded form.
Then\ttt{ x="$\beta$" }says that the variable\ttt{ x }has the value $\alpha$.
Therefore, the last part of G\"odel's sentence says that the value of\ttt{ x
}is the following string, with\ttt{ \Q{x}{y} }and\ttt{ \Pvble{x"\bs1"y"\bs1"}
}replaced by the strings they stand for:
\begin{quote}
\ttt{Ex:Ey:~\Q{x}{y} \& \tl\Pvble{x"\bs1"y"\bs1"} \& x=}
\end{quote}
This and\ttt{ \Q{x}{y} }together say that the value of\ttt{ y }is the
following string, with\ttt{ \Qp{x}{y} }and\\
\ttt{ \Pvblep{x"\bs1"y"\bs1"} }replaced by the strings they stand for:
\begin{quote}
\ttt{Ex:Ey:~\Qp{x}{y} \& \tl\Pvblep{x"\bs1"y"\bs1"} \& x=}
\end{quote}

The string literal\ttt{ "\bs1" }denotes the string\ttt{ " }.
Thus the value of\ttt{ x"\bs1"y"\bs1" }is the value of\ttt{ x }followed
by\ttt{ " }followed by the value of\ttt{ y }followed by\ttt{ " }.
Remembering that spaces and division to lines are only for simplifying
reading and not part of the real string, we see that the value of\ttt{
x"\bs1"y"\bs1" }is G\"odel's sentence.
Furthermore,\ttt{ \tl\Pvble{x"\bs1"y"\bs1"} }claims that\ttt{ x"\bs1"y"\bs1"
}is not provable.
To summarize, the other parts of G\"odel's sentence make\ttt{ x"\bs1"y"\bs1"
}be G\"odel's sentence, and\ttt{ \tl\Pvble{x"\bs1"y"\bs1"} }says that it is
not provable.
Altogether, G\"odel's sentence claims that G\"odel's sentence is not provable.

The formula\ttt{ \Pvble{} }specifies a proof system for strings.
G\"odel's sentence is not a single sentence, instead, each proof system for
strings has its own\ttt{ \Pvble{} }and thus its own G\"odel's sentence.
G\"odel's sentence of a proof system for strings claims that G\"odel's
sentence of that system is not provable in that system.

The Turing machine that halts immediately independently of the input
represents a proof system for strings that proves every sentence.
This proof system is useless, because for any sentence that it proves, it also
proves its negation.
So it proves many false sentences.
However, it serves as an example of a proof system that proves its own
G\"odel's sentence.

Consider now any proof system for strings that proves its own G\"odel's
sentence.
Because the sentence claims that the system does not prove it, the system has
proven a false sentence.
Consider then any proof system for strings that does not prove its own
G\"odel's sentence.
Its G\"odel's sentence thus expresses a true claim, and is thus a true
sentence that the system does not prove.

We have proven the following.

\begin{theorem}
Each recursively enumerable proof system for the first-order language on
finite character strings with string literals, concatenation, and $=$, either
proves a false sentence or fails to prove a true sentence.
\end{theorem}
That is, there is no recursively enumerable proof system for strings that
proves precisely the true sentences and nothing else.
No recursively enumerable proof system for strings can precisely capture the
true claims on strings that can be expressed in our language.
This is the incompleteness theorem for strings.

\section{Incompleteness of Natural Number Arithmetic}\label{S:arithmetic}

In this section we show that natural number arithmetic can simulate strings
and their concatenation, and conclude that also natural number arithmetic is
incomplete.

Our language on natural number arithmetic uses the same characters as our
language on strings in Section~\ref{S:language}.
A \emph{number literal} is either\ttt{ 0 }or any non-empty finite sequence of
digits that does not start with\ttt{ 0 }.
A \emph{variable} is any string that starts with a lower case letter and then
consists of zero or more digits.
The value of a variable is a natural number.
A \emph{term} is a variable, a number literal, or any of the following, where
$t$ and $u$ are terms:\ttt{~($t$) },\ttt{ $t$+$u$ }, or\ttt{ $t$*$u$ }.
The parentheses are used in the familiar way,\ttt{ + }denotes addition,
and\ttt{ * }denotes multiplication.
Furthermore,\ttt{ * }has higher precedence than\ttt{ +~}, that is,\ttt{
$t$+$u$*$v$ }denotes the same as\ttt{ $t$+($u$*$v$) }.
Atomic propositions and formulae are defined like in Section~\ref{S:language}.

\newcommand{\num}{\mathit{num}}
We now introduce a one-to-one correspondence between strings and natural
numbers.
Let $p=53$, and let the $53$ characters in the character set be given numbers
from $1$ to $53$.
If $c'_i$ is a character, then let its number be denoted with $c_i$.
The string $c'_1 c'_2 \cdots c'_n$ has the number
$$\num(c'_1 c'_2 \cdots c'_n)\quad = \quad c_1p^{n-1} + c_2p^{n-2} + \ldots +
c_{n-2}p^2 + c_{n-1}p + c_n\quad\textrm{.}$$
So the empty string has the number $0$, and the number of any string
consisting of precisely one character is the number of that character.
Let $\iota_n$ denote the string of length $n$ whose every character has number
$1$.
We have $\num(\iota_n) = p^{n-1} + \ldots + p + 1$, and $\iota_0$ is the empty
string.

We have to show that this mapping is indeed one-to-one.
To do that, for each string $c'_1 c'_2 \cdots c'_n$ we introduce a
\emph{successor} and prove that the number of the successor is always one
bigger than the number of the string itself.
If $c_i = p$ for every $1 \leq i \leq n$, then the successor is defined as
$\iota_{n+1}$.
We have
\begin{quote}
\begin{tabular}{l@{~}l}
& $\num(\iota_{n+1})$\\
$-$ & $\num(c'_1 c'_2 \cdots c'_n)$
\end{tabular}
$=$
\begin{tabular}{r@{~}r@{~}r@{~}r@{~}r@{~}r@{~}r@{~}r}
& $p^n$ & $+$ & $p^{n-1} $ & $+\ \ldots\ +$ & $p$ & $+$ & $1$\\
$-$ & $pp^{n-1}$ & $-$ & $pp^{n-2}$ & $-\ \ldots\ -$ & $p \cdot 1$
\end{tabular}
$=\ 1$ .
\end{quote}
In the opposite case, at least one of $c_1$, \ldots, $c_n$ is not $p$.
Let $j$ be the last such index, that is, $1 \leq j \leq n$, $c_j \neq p$, and
$c_i = p$ when $j < i \leq n$.
The successor is defined as the string $d'_1 d'_2 \cdots d'_n$, where $d_i =
c_i$ when $1 \leq i < j$, $d_j = c_j+1$, and $d_i = 1$ when $j < i \leq n$.
We have
\begin{quote}
$\num(d'_1 d'_2 \cdots d'_n) - \num(c'_1 c'_2 \cdots c'_n)$ $=$\\
\mb\hfill\begin{tabular}
{r@{~}r@{~}r@{~}r@{~}r@{~}r@{~}r@{~}r@{~}r@{~}r@{~}r@{~}r@{~}r@{~}r}
& $c_1 p^{n-1}$ & $+\ \ldots\ +$ & $c_{j-1} p^{n-j+1} $ & $+$ &
$(c_j+1)p^{n-j}$ &&& $+$ & $p^{n-j-1}$
& $+\ \ldots\ +$ & $p$ & $+$ & $1$\\
$-$ & $c_1 p^{n-1}$ & $-\ \ldots\ -$ & $c_{j-1} p^{n-j+1}$ & $-$ &
$c_j p^{n-j}$ & $-$ & $pp^{n-j-1}$ & $-$ & $pp^{n-j-2}$
& $-\ \ldots\ -$ & $p \cdot 1$
\end{tabular}\\
$=\ 1$ .
\end{quote}

We see that the empty string, its successor, the successor of that string, and
so on are in one-to-one correspondence with the natural numbers $0$, $1$, $2$,
and so on.
It remains to be proven that this sequence of strings covers all strings.
It does not contain any string twice, because the corresponding natural
numbers are all distinct.
So it contains infinitely many distinct strings.
For any $n$, there is only a finite number of strings of length $n$.
So the sequence cannot get stuck at any length $n$.
The only case where the successor is of different length than the string
itself is when the successor is $\iota_{n+1}$.
So the sequence covers at least $\iota_0$, $\iota_1$, $\iota_2$, and so on.
Between $\iota_n$ and $\iota_{n+1}$, including $\iota_n$ but not
$\iota_{n+1}$, the sequence goes through $\num(\iota_{n+1}) - \num(\iota_n) =
p^n$ strings of length $n$.
The number of strings of length $n$ is $p^n$, so the sequence goes through all
of them.

We have shown that our correspondence between strings and natural numbers is
one-to-one.

Our next task is to represent concatenation of strings as a formula on their
numbers.
The definition of $\num$ yields immediately
$$\num(c'_1 \cdots c'_n d'_1 \cdots d'_m) \ =\ p^m\num(c'_1 \cdots c'_n) +
\num(d'_1 \cdots d'_m)\quad\textrm{.}$$
To present this in our language, we have to extract $p^m$ from $\num(d'_1
\cdots d'_m)$ only using the language.
Let $y = \num(d'_1 \cdots d'_m)$.
We have $\num(\iota_m) \leq y < \num(\iota_{m+1})$, that is, $p^{m-1} + \ldots
+ 1 \leq y < p^m + \ldots + 1$.
Multiplying this by $p-1$ we get $p^m - 1 \leq y(p-1) < p^{m+1} - 1$, to which
adding $y+1$ yields $p^m + y \leq yp+1 < p^{m+1} + y$.
If $m' > m$, then $p^{m'} + y \leq yp+1$ does not hold, and if $m' < m$, then
$yp+1 < p^{m'+1} + y$ does not hold.
Therefore, $k = p^m$ if and only if $k$ is a power of $p$ and $k + y \leq yp+1
< pk + y$.

A prime number is a natural number greater than $1$ that cannot be represented
as a product of two natural numbers greater than $1$.
If $p$ is a prime number and $p^m = xy$, then, for some $0 \leq i \leq m$, $x
= p^i$ and $y = p^{m-i}$.
Therefore, and because $53$ is a prime number, the property that $k$ is a
power of $53$ can be formulated as follows.
\newcommand{\PowP}[1]{\ts{Pow53(}#1\ts{)}}
\begin{quote}
\ttt{\PowP{k} \defby\ ( Ax:Ay:~k=x*y -> x=1 | Ez:~x=53*z )}
\end{quote}
That $x<y$ can be expressed as follows.
\newcommand{\less}[2]{\ts{lt(}#1\ts{, }#2\ts{)}}
\begin{quote}
\ttt{\less{x}{y} \defby\ ( Ei:~y=x+i+1 )}
\end{quote}
Based on these considerations, if\ttt{ x }and\ttt{ y }are the numbers of two
strings, then the number of the concatenation of the strings is obtained as
follows.
\newcommand{\Cat}[3]{\ts{Cat(}#1\ts{, }#2\ts{, }#3\ts{)}}
\begin{quote}
\ttt{\Cat{x}{y}{z} \defby\\
\mb\hspace{1cm}( Ek:~\PowP{k} \& \less{y*53+1}{53*k+y} \&
\tl\less{y*53+1}{k+y} \& z=k*x+y )}
\end{quote}
Atomic propositions in our language on strings are of the form\ttt{ $t_1
\cdots t_m$=$u_1 \cdots u_n$ }or\ttt{ $t_1 \cdots t_m$\nq$u_1 \cdots u_n$~},
where $t_1$, \ldots, $t_m$, $u_1$, \ldots, $u_n$ are variables or string
literals.
They can be replaced as shown below for\ttt{ = }, where\ttt{ t }and\ttt{ u
}are two variable names that are different from the $t_i$ and $u_j$.

\begin{quote}
\ttt{( Et:Eu:~t=u\\
\mb\hspace{1cm}\& ( Eu:~\Cat{u}{$t_m$}{t} \& Et:~\Cat{t}{$t_{m-1}$}{u} \&
Eu:~\Cat{u}{$t_{m-2}$}{t} \& $\ldots$ )\\
\mb\hspace{1cm}\& ( Et:~\Cat{t}{$u_n$}{u} \& Eu:~\Cat{u}{$u_{n-1}$}{t} \&
Et:~\Cat{t}{$u_{n-2}$}{u} \& $\ldots$ )\\
)}
\end{quote}

There is a Turing machine $T_1$ that inputs a sentence in the language on
strings, replaces each string literal by its number, and replaces each atomic
proposition as shown above.
(We could have made this easier for the Turing machine but harder for the
reader by, in Section~\ref{S:language}, not allowing more than one character
in any string literal, not allowing more than one variable and/or string
literal in a term, and instead declaring that \ttt{<x+y:z>} expresses that
\ttt{xy=z}.)
If there is a Turing machine $T_2$ that halts on the true sentences in the
language on natural numbers and fails to halt on false sentences, then there
is a Turing machine $T$ that first runs $T_1$ and then runs $T_2$ on the
result.
By construction, $T$ halts if and only if its input string is a true sentence
in the language on strings.
But we proved in Section~\ref{S:strings} that such a Turing machine does not
exist.
Therefore, $T_2$ does not exist.
We have proven the following.

\begin{theorem}
Each recursively enumerable proof system for the first-order language on
natural numbers with $0$, $1$, $+$, $\cdot$, and $=$, either proves a false
sentence or fails to prove a true sentence.
\end{theorem}

\section{Versions of G\"odel's First Incompleteness
Theorem}\label{S:versions}

We proved that any recursively enumerable theory of natural numbers with zero,
one, addition, multiplication, equality, logical and, logical not, and the
universal quantifier either proves a false sentence or fails to prove a true
sentence.
Although this theorem is widely called G\"odel's first incompleteness theorem,
it falls short of what G\"odel presented in~\cite{God31}.
It assumes that the truth or falsehood of a sentence can be reasonably talked
about, even if the theory does not prove either.
(When we say that a theory proves a sentence false, we mean that the theory
proves the negation of the sentence.)
This assumption has been criticized.
Perhaps for this reason, G\"odel went beyond this version.
To discuss this, we first make the following observation.

In the presence of truth and falsehood as we usually consider them, a sentence
and its negation cannot both be true.
Furthermore, for each sentence, either it or its negation is true.
A theory is \emph{consistent} if and only if in no case it proves both a
sentence and its negation.
A theory is \emph{complete} if and only if in each case it proves the sentence
or its negation.
(The word ``complete'' is used in more than one meaning in mathematical logic.
This is the meaning we use here.)
Therefore, if a theory proves only true sentences and proves all of them, then
it is consistent and complete.

The notions of consistency and completeness do not rely on a pre-defined
notion of truth of a sentence.
However, they do not together mean the same as ``proves only true sentences
and proves all of them'', because it may be that the theory fails to prove a
true sentence and instead proves its false negation.
Indeed, there are consistent and complete theories whose language is the same
as the language of natural number arithmetic.
An example is obtained by letting $0$ denote \ts{false}, $1$ denote \ts{true},
$+$ denote $\vee$, and $\cdot$ denote $\wedge$, and by adopting the usual
axioms and inference rules of $=$ and propositional logic together with two
special rules: ``$\forall x: P(x)$ is equivalent to $P(0) \cdot P(1)$'' and
``$\exists x: P(x)$ is equivalent to $P(0) + P(1)$''.
This theory could well be called a first-order theory of truth values.
It proves sentences that are false from the point of view of natural number
arithmetic, such as $1+1 = 1$ (which represents $\ts{true} \vee \ts{true} =
\ts{true}$).
It is a consistent and complete theory, but a wrong theory for natural number
arithmetic although it has the same language.

Another way to look at this is that the replacement of the notions of truth
and falsehood by completeness and consistency disconnect the language from
natural numbers, leaving only two uninterpreted constant symbols $0$ and $1$,
and two uninterpreted binary operator symbols $+$ and $\cdot$.
The mere fact that the symbols look familiar does not give them any formal
properties.
Instead, to make them again have a link with natural number arithmetic, some
axioms and inference rules are needed.

In conclusion, the right liberation of the incompleteness result from the
notion of pre-defined truth is that no ``sufficiently strong'' theory of
natural number arithmetic is consistent and complete.
Here ``sufficient strength'' has two aspects.
First, the notion of first-order theories has a standard set of logical axioms
and inference rules.
It is assumed.
Second, enough properties of natural numbers are assumed in the form of
axioms, to ensure that the theory indeed is a theory of natural numbers
instead of, say, the theory of truth values sketched above.
Not much is needed.
A rather weak axiom system known as \emph{Robinson arithmetic}
suffices~\cite{Rob50,TMR53}.
It is otherwise the same as the well-known Peano arithmetic, but the induction
axiom has been replaced by the axiom ``each natural number is either $0$ or
the result of adding $1$ to some natural number.''

G\"odel did not prove the theorem in the form stated above.
Instead of consistency, he used the stronger notion called
\emph{$\omega$-consistency}.
A theory is not $\omega$-consistent if and only if it is not consistent or,
for some formula $P$ with one free variable $x$, it proves both $\exists x:
P(x)$ and each one of $\neg P(0)$, $\neg P(1)$, $\neg P(2)$, \ldots.
That a theory proves each one of $\neg P(0)$, $\neg P(1)$, $\neg P(2)$,
\ldots\ does not necessarily imply that it proves their conjunction $\forall
x: \neg P(x)$, because no proof can go through an infinite number of
cases one by one (proofs must be finite), and a common pattern that would
facilitate proving them simultaneously in a single proof does not necessarily
exist.
Even so, intuition says that if none of $P(0)$, $P(1)$, $P(2)$, and so on
holds, then there is no $x$ such that $P(x)$ holds, that is, $\neg \exists x:
P(x)$ holds.
So a healthy theory of natural number arithmetic must be not only consistent,
but also $\omega$-consistent.

\newcommand{\prf}{\mathsf{Prf}}
G\"odel's result was that such a theory cannot be complete.
Let $\prf(x,y)$ denote the claim that natural number $x$ is the encoding of a
proof of the sentence encoded by natural number $y$.
This claim can be formulated in natural number arithmetic.
Furthermore, if $\prf(x,y)$ holds, then $\prf(x,y)$ can be proven, and if
$\neg\prf(x,y)$ holds, then $\neg\prf(x,y)$ can be proven.
Together with the requirement of ``effectivity'', this implies that the proof
system must be recursive in the sense of Section~\ref{S:background}.
That is, there is a mechanical test which, for any string, tells whether it is
a valid proof, where also the answer ``no'' is given explicitly instead of
just never answering anything.
Although this assumption is strictly stronger than recursive enumerability,
proof systems typically satisfy it.

The proof system must also facilitate the simple reasoning steps in the
sequel.

G\"odel's self-referential sentence is $\neg\exists x: \prf(x,g)$, where $g$
is its own encoding as a natural number.
Assume first that the proof system proves $\neg\exists x: \prf(x,g)$.
Then there is a natural number $p$ that is the encoding of some proof of
$\neg\exists x: \prf(x,g)$.
By the strength assumption above, the proof system proves $\prf(p,g)$.
From it the proof system can conclude $\exists x: \prf(x,g)$.
So it proves both $\exists x: \prf(x,g)$ and its negation, and is thus not
consistent.
The case remains where the proof system does not prove $\neg\exists x:
\prf(x,g)$.
Then no natural number is the encoding of a proof of $\neg\exists x:
\prf(x,g)$.
By the strength assumption above, the proof system proves $\neg \prf(0,g)$,
$\neg \prf(1,g)$, and so on.
If the proof system is $\omega$-consistent, then it does not prove $\exists x:
\prf(x,g)$.
So it leaves both $\exists x: \prf(x,g)$ and $\neg\exists x: \prf(x,g)$
without a proof, and is thus incomplete.

Later Rosser found a modification to the proof that allows to replace
$\omega$-consistency with consistency~\cite{Ros36}.
We call his self-referential sentence R.
It is $\forall x: (\prf(x,r) \rightarrow \exists y: y \leq x \wedge
\prf(y,\bar r))$, where $r$ is the encoding of R and $\bar r$ is the encoding
of $\neg$R.
If $p$ is the encoding of a proof of R, then the system proves R, $\prf(p,r)$,
and $\exists y: y \leq p \wedge \prf(y,\bar r)$.
If such an $y$ indeed exists, then the proof whose encoding is $y$ yields
$\neg$R, so the system proves a contradiction.
Otherwise, the system proves $\neg\prf(0,\bar r) \wedge \neg\prf(1,\bar r)
\wedge \cdots \wedge \neg\prf(p,\bar r)$, yielding $\neg\exists y: y \leq p
\wedge \prf(y,\bar r)$, a contradiction again.
$\omega$-con\-sistency is not needed, because $\neg\prf(0,\bar r) \wedge
\cdots \wedge \neg\prf(p,\bar r)$ is a finite expression and thus a sentence.

If $p$ is the encoding of a proof of $\neg$R, then the system proves
$\prf(p,\bar r)$ and $\exists x: (\prf(x,r) \wedge \neg\exists y: y \leq x
\wedge \prf(y,\bar r))$, yielding $\exists x: \prf(x,r) \wedge \neg(p \leq
x)$.
Like above, the system proves one or another contradiction, depending on
whether any of $0$, $1$, \ldots, $p-1$ is the encoding of a proof of R.

In conclusion, if the system is consistent, then it proves neither R nor
$\neg$R, and is thus incomplete.

The above proof of Rosser's theorem uses the symbol $\leq$ that is not part of
the first-order language on arithmetic.
It can be expressed as mentioned in Section~\ref{S:background}.
The crucial property is that if $c$ is a natural number constant, then no
other natural numbers than $0$, $1$, \ldots, $c$ have the properties that $x
\leq c$ and $\neg(c+1 \leq x)$.
This can be proven from Peano arithmetic, but in the case of other axiom
systems, specific axioms on $\leq$ may be needed.

In his original publication~\cite{God31} G\"odel also sketched a proof of a corollary that is now known as G\"odel's second incompleteness theorem.
It says that natural number arithmetic does not prove its own consistency, if
it indeed is consistent.
What is more, no recursive consistent theory that contains natural number
arithmetic proves its own consistency.
The significance of this result is the following.
Some mathematical principles are easy to accept, while some others have raised
doubts.
The questionable principles would become more acceptable, if, with a proof
that only uses easily acceptable principles, they were proven to not yield
contradictions.
G\"odel's second incompleteness theorem rules out perhaps not all, but at
least the most obvious approaches to such proofs.

\section{Related Work and Conclusions}\label{S:conclusions}

We have shown that any recursively enumerable first-order theory of finite
character strings with concatenation and equality either proves a false
sentence or fails to prove a true sentence.
From this we derived a similar result about natural number arithmetic with
addition, multiplication, and equality, obtaining the ``either proves a false
sentence or fails to prove a true sentence'' version of G\"odel's first
incompleteness theorem.

A \emph{halting tester} is a Turing machine that reads any Turing machine $M$
together with its input $I$ and tells whether $M$ halts, if executed on $I$.
In addition to inventing his machines, Alan Turing proved that there is no
halting tester~\cite{Tur36}.
(The modern version of this proof is very simple.)

Our proof of G\"odel's theorem is based on encoding each claim of the form
``$M$ halts on $I$'' as a sentence in natural number arithmetic.
It is not the first such proof.
If each encoded sentence ``$M$ halts on $I$'' or its negation were provable by
a system that only proves true sentences, then a halting tester would be
obtained by letting a Turing machine test all finite character strings until
it finds a proof of halting or non-halting.
Therefore, the system has an unprovable true sentence.
Essentially the same reasoning can be expressed in another words by pointing
out that the set of provable sentences is recursively enumerable, but the set
of true sentences is not recursively enumerable, because the true sentences of
the form ``$M$ does not halt on $I$'' cannot be enumerated.
So truth and provability do not match.
This proof is given in, e.g.,~\cite[p.~354]{HoU79} (leaving the (A) mentioned
below as a doubly starred exercise!) and~\cite[pp.~288--291]{Koz97}, underlies
the proof in~\cite[p.~134]{Pap94}, and is at least hinted at
in~\cite[p.~64]{Kle43}.

Two major difficult technicalities in G\"odel's proof are (A) to show that
reasoning or computation can be encoded as properties of natural numbers
(so-called \emph{G\"odel numbering}), and (B) to give a formula access to its
own number.
The proof based on non-existence of halting testers makes (B) trivial.
It makes it necessary to check or believe that, given $M$ and $I$, a Turing
machine can perform the construction in (A).
Fortunately, it is rather obvious.

Because natural number arithmetic has no direct construct for expressing
finite sequences of natural numbers, (A) is surprisingly difficult.
To do (A), often the Chinese Remainder Theorem is used.
That brings discussion so far from the main topic that some expositions simply
skip the issue.
Alternatively, one may add the exponentiation operator $n^m$ to the theory, as
was done in~\cite[p.~135]{Pap94}.
In~\cite{Koz97}, Dexter Kozen made (A) relatively easy by using a slightly
less straighforward representation for halting computations than we did in
Section~\ref{S:machine}, and treating natural numbers essentially as finite
sequences of $p$-ary digits, where $p$ is a prime.
Numbers that were known to be powers of $p$, but not known which power, were
used to extract individual digits.
Thanks to padding with blanks, the representation of each configuration during
a computation used the same number of digits.
As a consequence, there was a number $c$ such that if $y$ extracts a digit in
a configuration, then $yc$ extracts the corresponding digit in the next
configuration.

Our proof made (A) and (B) easy by doing them in a formalism that is very
amenable to them.
The most advanced number-theoretic property needed in the whole proof is that
if a number is a power of prime $p$, then all its factors other than $1$ are
divisible by $p$.
The Chinese Remainder Theorem was not used and the exponentiation operator was
not added to the language.
Turing machines were referred to twice: as the basis of the definition of
``recursively enumerable proof system'', and as devices that can perform a
simple syntactic transformation.

It seems obvious that the incompleteness of theories of finite character
strings can also be proven with the approach in~\cite{HoU79,Koz97,Pap94}.
Then one may continue like in Section~\ref{S:arithmetic}.
In this combined approach, the formula\ttt{ \Q{x}{y} }would not be needed (but
the computability of (A) by a Turing machine would).

Neil D.\ Jones has proven the incompleteness of first-order theories of nested
lists with concatenation~\cite[p.\ 202]{Jon97}.
Also this proof is based on the non-existence of halting testers.
The counterpart of (A) is trivial, because the formalism supports it directly.
Nested lists are a strong formalism that can easily express natural numbers,
so this result is not surprising.

Finite character strings with concatenation may at first sight seem a poor
formalism: a data type with infinitely many distinct values could not be much
simpler.
On the other hand, all computation reduces to the manipulation of finite
character strings both in theory (Turing machines) and in practice (files are
finite sequences of bytes).
So the incompleteness of first-order theories of finite character strings with
concatenation seems too obvious to be a new result.
To prove it, it suffices to cite G\"odel and then show that strings can
simulate arithmetic.
Such simulations have been studied at least in~\cite{CFM74,Qui46}.
However, neither publication explicitly mentions the incompleteness of strings
and, indeed, the author has failed to find any mention of it in the
literature.
What is more, in this paper the proof was simplified by simulating in the
opposite direction, that is, by proving the incompleteness of strings directly
and then deriving the incompleteness of arithmetic as a corollary.
This idea seems to be new.

Even if it turns out that the approach of this paper is not novel, we hope
that the paper helps the readers understand G\"odel's famous result.

\paragraph{Acknowledgements.}
This version of the paper has benefited from the good comments given by the
anonymous reviewers.

\bibliography{AFLgodel}
\bibliographystyle{eptcs}

\end{document}